\documentstyle[11pt]{article}

\newlength{\minitwocolumn}
\catcode`\@=11
\def\relaxnext@{\let\next\relax}
\font\tenmsy=msym10 scaled\magstep1
\font\sevenmsy=msym7 scaled\magstep1
\font\fivemsy=msym5  scaled\magstep1
\newfam\msyfam
\textfont\msyfam=\tenmsy
\scriptfont\msyfam=\sevenmsy
\scriptscriptfont\msyfam=\fivemsy

\newtheorem{thm}{Theorem}[section]
\newtheorem{prop}[thm]{Proposition}
\newtheorem{lem}[thm]{Lemma}
\newtheorem{cor}[thm]{Corollary}

\newcommand{\bra}{_{\epsilon}
\langle \Psi_N(\lambda_1,\cdots,\lambda_N)\vert}
\newcommand{\ket}{\vert \Psi_N(\lambda_1,\cdots,\lambda_N)
\rangle_{\epsilon}}

\begin{document}
\begin{flushright}
December 1997
\end{flushright}
\vspace{44pt}
\begin{center}
\begin{Large}
{\bf Dynamical Correlation Functions for an Impenetrable Bose Gas

with Open Boundary Conditions.}
\end{Large}

\vspace{35pt}
By
\vspace{10pt}

Takeo Kojima
\vspace{20pt}

{\it Research Institute for Mathematical Sciences,
     Kyoto University, Kyoto 606, Japan}

\vspace{60pt}

\underline{Abstract}

\end{center}

We study the time and temperature dependent correlation functions
for an impenetrable bose gas with open boundary conditions.
We derive the Fredholm determinant formulae for the correlation
functions, by means of the Bethe Ansatz.
In the case of time independent ground state,
our Fredholm determinant 
formulae degenerate to the one which have been obtained by the help
of fermions \cite{K}.

\vspace{25pt}

\newpage
\section{Introduction}
In the standard treatment of quantum integrable models,
one starts with a finite box and impose periodic boundary conditions,
in order to ensure integrability.
Recently, there has been increasing interest in exploring
other possible boundary conditions compatible with integrability.
These other possible boundary conditions are called 
``open boundary conditions''.

With open boundary conditions, the works on the two dimensional 
Ising model are among the earliest.
By the help of graph theoretical approach, B.M. McCoy and T.T. Wu \cite{M.W.}
studied the two dimensional Ising model with open boundary conditions.
They calculated the local magnetizations.
E.K. Sklyanin \cite{S} 
began the Bethe Ansatz approach to open boundary problems.
M. Jimbo et.al. \cite{J} calculated correlation functions of local operators
for the antiferromagnetic XXZ chains with open boundary conditions,
using Sklyanin's algebraic Bethe Ansatz framework and
representation theoretical approach invented by Kyoto school
\cite{D.F.J.M.N.} \cite{J.M.}.
T. Kojima \cite{K} studied 
the ground state correlation functions for
an impenetrable Bose gas with open boundary conditions :
\begin{eqnarray}
\langle \psi(x_1)\psi^\dagger(x_2)\rangle.
\end{eqnarray}
He derived the Fredholm minor determinant representations for 
the ground state correlation functions by the help of fermions,
which have the integral kernel :
\begin{eqnarray}
\frac{\sin(\lambda-\mu)}{\lambda-\mu}\pm
\frac{\sin(\lambda+\mu)}{\lambda+\mu}.\label{kernel:deg}
\end{eqnarray}
The integral intervals depend on the space parameter $x_1, x_2$.
In this paper we study an impenetrable Bose gas with open boundary conditions.
We are interested in the finite-temperature dynamical correlation functions :
\begin{eqnarray}
\langle \psi(x_1,0)\psi^\dagger(x_2,t)\rangle_{T}.
\end{eqnarray}
We derive the Fredholm determinant representations for
the dynamical correlation functions by the coordinate Bethe Ansatz,
which have the integral kernel :
\begin{eqnarray}
\frac{L(\lambda,\mu)}{\lambda-\mu}\pm
\frac{L(\lambda,-\mu)}{\lambda+\mu},
\end{eqnarray}
where we have used
\begin{eqnarray}
L(\lambda,\mu)&=&e^{it\lambda^2}\sin(x_1(\lambda-\mu))+
e^{it\mu^2}\sin(x_2(\lambda-\mu))\\
&~&
+\frac{2}{\pi}\int_{-\infty}^{\infty}
\left(\frac{1}{s-\mu}-\frac{1}{s-\lambda}\right)e^{its^2}
\sin((s-\mu)x_1)\sin((s-\lambda)x_2)ds.
\nonumber
\end{eqnarray}
The mesure of integral depends on the temperature parameter $T\geq0$.
For the special case that time $t=0$ and temperature $T=0$,
our formulas degenerate to the one which has been obtained \cite{K}.
In the case of periodic bounadary conditions,
V.E. Korepin and N.A. Slavnov \cite{K.S.} 
calculated the dynamical correlation functions
for an impenetrable Bose gas by the corrdinate Bethe Ansatz.

Now a few words about the organization of the paper.
In section 2 we formulate the problem.
In section 3 we obtain the determinant formulae for
the field form factors.
In section 4 we obtain the Fredholm determinant representation
for the dynamical correlation functions.
In section 5 we consider the special case that time $t=0$
and 
derive the Fredholm minor determinant representations
for the finite-temperature fields correlation functions.
We show that our Fredholm formulae degenerate to the one which
has been obtained \cite{K}.

\section{Formulation}
The purpose of this section is to formulate the problem.
The Hamiltonian of our model is given by
\begin{eqnarray}
{\it H}=\int_0^L dx
\left(\partial_x\psi^\dagger\partial_x\psi
+c\psi^\dagger \psi^\dagger \psi \psi-h \psi^\dagger \psi\right)
+h_0\left(\psi^\dagger(0)\psi(0)-\psi^\dagger(L)\psi(L)\right).
\end{eqnarray}
Here the fields $\psi(x)$ and $\psi^\dagger(x)~(x \in {\bf R})$
are canonical Bose fields
given by
\begin{eqnarray}
[\psi(x),\psi^\dagger(y)]=\delta(x-y),~~
[\psi(x),\psi(y)]=[\psi^\dagger(x),\psi^\dagger(y)]=0,~~(x,y \in {\bf R}),
\end{eqnarray}
and $L>0$ is the size of box. The parameters $h>0$ and $h_0 \in {\bf R}$
represent
the chemical potential and the boundary chemical potential respectively.
We only consider the case of the coupling constant $c=\infty$, so-called
``impenetrable case''.
The Hamiltonian $\it H$ acts on the Fock space of the Bose fields
defined by the following relations between the Fock vacuum $\vert 0 \rangle$
and the Bose fields :
\begin{eqnarray}
\langle 0 \vert \psi^{\dagger}(x)=0,~\psi(x)\vert 0 \rangle=0,~
\langle 0 \vert 0 \rangle=1.
\end{eqnarray}
A $N$-particle state vector $\vert \Psi_N\rangle$ is given by
\begin{eqnarray}
\vert \Psi_N\rangle
=\int_0^Ldz_1\cdots\int_0^Ldz_N\psi_N(z_1,\cdots,z_N)
\psi^\dagger(z_1)\cdots \psi^\dagger(z_N)\vert 0 \rangle,
\end{eqnarray}
where the integrand $\psi_N(z_1,\cdots,z_N)$ is a $\bf C$-valued function.
The eigenvector problem :
$
{\it H}\vert \Psi_N \rangle
=E_N \vert \Psi_N \rangle,~(E_N \in {\bf R})
$,
is equivalent to the quantum mechanics problem defined by the following
four conditions of the integrand function $\psi_N(z_1,\cdots,z_N)$.

\begin{enumerate}
\item
The wave function $\psi_N=\psi_N(z_1,\cdots,z_N)$ satisfies
the free-particle Schr\"odinger equation in the case of variables
$0 < z_i \neq z_j < L$ :
\begin{eqnarray}
-\sum_{j=1}^N \left(\frac{\partial}{\partial z_j}\right)^{2}
\psi_N(z_1,\cdots,z_N)=
E_N\cdot\psi_N(z_1,\cdots,z_N),~~(0< z_i\neq z_j < L,~E_N \in {\bf R}).
\end{eqnarray}

\item
The wave function $\psi_N$ is symmetric with respect to the variables :
\begin{eqnarray}
\psi_N(z_1,\cdots,z_N)=
\psi_N(z_{\sigma(1)},\cdots,z_{\sigma(N)}),~~(\sigma \in S_N).
\end{eqnarray}

\item
The wave function $\psi_N$ satisfies the integrable open boundary conditions :
\begin{eqnarray}
\left.\left(\frac{\partial}{\partial z_j}-h_0\right) \psi_N\right|_{z_j=0}=0,~~
\left.\left(\frac{\partial}{\partial z_j}+h_0\right) \psi_N\right|_{z_j=L}=0,~~(j=1,\cdots,N).
\end{eqnarray}
\item
The wave function $\psi_N$ vanishes whenever the corrdinates coincide :
\begin{eqnarray}
\left.\psi_N(z_1,\cdots,z_i,\cdots,z_j,\cdots,z_N)\right|
_{z_i=z_j}=0.
\end{eqnarray}
\end{enumerate}

The wave functions $\psi_N$ which satisfy the above four conditions were 
constructed \cite{K}. They are parametrized by the spectral parameters.

\begin{eqnarray}
&&\psi_N(z_1,\cdots,z_N\vert\lambda_1,\cdots,\lambda_N)\nonumber \\
&=&Cons.\prod_{1\leq j < k \leq N}{\rm sgn}(z_j-z_k)
\det_{1 \leq j,k \leq N}
\left(\lambda_j \cos(\lambda_j z_k)+h_0 \sin(\lambda_j z_k)\right).
\end{eqnarray}
Here the function ${\rm sgn}(x)=\frac{x}{\vert x \vert}$
and the spectral parameters $0\leq
\lambda_1<\lambda_2<\cdots<\lambda_N$ are determined by
the so-called Bethe Ansatz equations :
\begin{eqnarray}
\lambda_j=\frac{\pi}{L}I_j,~~(I_j \in {\bf N},~j=1,2,\cdots,N).\label{BAE}
\end{eqnarray}
The constant factor ``$Cons.$'' is deteminend by
\begin{eqnarray}
\langle \Psi_N(\lambda_1,\cdots,\lambda_N)\vert
\Psi_N(\lambda_1,\cdots,\lambda_N)\rangle = (2L)^N.
\end{eqnarray}
The eigenvalue $E_N(\{\lambda\})$ :
\begin{eqnarray}
{\it H}\ket = E_N(\{\lambda\})\ket,
\end{eqnarray}
is given by
\begin{eqnarray}
E_N(\{\lambda\})=\sum_{j=1}^N(\lambda_j^2-h).
\end{eqnarray}
We assume that the set
$
\{ \vert \Psi_N(\lambda_1,\cdots,\lambda_N)\rangle \}
_{{\rm all}\{\lambda\}_N ~ N \in {\bf N}}
$
is a basis of physical space of this model. Here the index
${\rm all}\{\lambda\}_N$
represents all the solutions of the Bethe Ansatz equations (\ref{BAE}).
This type assumption is usually called ``Bethe Ansatz''.
The following lemma is a foundation of our analysis.

\begin{lem}~~If the boundary condition $h_0$ takes the special
value $h_0=0, \infty$,
the eigenvectors $\vert \Psi(\{\lambda\})\rangle$
satisfy orthogonality relations.
\begin{eqnarray}
\langle \Psi_N(\lambda_1,\cdots,\lambda_N)\vert
\Psi_N(\mu_1,\cdots,\mu_N)\rangle =(2L)^N 
\prod_{j=1}^N \delta_{\lambda_j,\mu_j},~~
(h_0=0,\infty).\label{orthogonal}
\end{eqnarray}
Here $\delta_{\lambda,\mu}$ is Kronecker Delta.
\end{lem}
To prove the above lemma, we have used the Bethe-Ansatz equations of 
the spectral parameters.
In the sequel we use the orthogonality relations of the eigenstates,
therefore we concentrate our attentions to the case of the special bounadary
conditions : $h_0=0,\infty$.
The boundary conditions $h_0=0$ and $h_0=\infty$ are called
Neumann, Dirichlet, respectively.
In the sequel we use the following abberivations.
\begin{eqnarray}
\vert \Psi_N(\lambda_1,\cdots,\lambda_N)\rangle_+ 
{\rm ~for~Neumann},~~~
\vert \Psi_N(\lambda_1,\cdots,\lambda_N)\rangle_- 
{\rm ~for~Dirichlet}.
\end{eqnarray}
The constant ``{\it Cons.}'' is given by
\begin{eqnarray}
Cons.=\left\{
\begin{array}{cc}
\frac{2^N}{\sqrt{(1+\delta_{\lambda_1,0})N!}}
\left(\prod_{j=1}^N\lambda_j\right)^{-1},
&{\rm for~Neumann},\\
\frac{1}{\sqrt{N!}}\left(\frac{2i}{h_0}\right)^N,&{\rm for~Dirichlet}.
\end{array}\right.
\end{eqnarray}
From the orthogonal relations $(\ref{orthogonal})$
and the so-called ``Bethe Ansatz'',
we arrive at the completeness relations.
\begin{eqnarray}
id=\sum_{N=0}^{\infty}\sum_{{\rm all}\{\lambda\}_N}
\frac{\vert \Psi_N(\lambda_1,\cdots,\lambda_N)\rangle_{\epsilon}
~_{\epsilon}\langle\Psi_N(\lambda_1,\cdots,\lambda_N)\vert}
{_{\epsilon}\langle\Psi_N(\lambda_1,\cdots,\lambda_N)\vert
\Psi_N(\lambda_1,\cdots,\lambda_N)\rangle_{\epsilon}}.
\label{completeness}
\end{eqnarray}
The Bose fields $\psi(x,t),\psi^\dagger(x,t)$
are developed by the time $t$ by
\begin{eqnarray}
i\partial_t \psi =[\psi, H],~~~
i\partial_t \psi^\dagger
=[\psi^\dagger, H].
\end{eqnarray}
More explicitly the time dependence of the Bose fields are written by
\begin{eqnarray}
\psi(x,t)=e^{iHt}\psi(x)e^{-iHt},~~~
\psi^\dagger(x,t)=e^{iHt}\psi^\dagger(x)e^{-iHt}.
\end{eqnarray}

In this paper we are interested in the dynamical correlation functions
$\langle \psi(x_1,t_1)\psi^\dagger(x_2,t_2)\rangle_{\epsilon,T}$
defined by the following way.
For the nonzero temperature $T>0$,
the dynamical correlation functions for $N$ state are defined by
the summation of the every states :
\begin{eqnarray}
&&\langle \psi(x_1,t_1)\psi^\dagger(x_2,t_2)\rangle_{\epsilon,N,T}=
\displaystyle \left\{\sum_{{\rm all} \{\lambda\}_N}
\exp\left(-\frac{E_N(\{\lambda \})}{T}\right)
\right\}^{-1}
\nonumber \\
&\times&\displaystyle \left\{\sum_{{\rm all} \{\lambda\}_N}
\exp\left(-\frac{E_N(\{\lambda \})}{T}\right)
\frac{\bra \psi(x_1,t_1)\psi^\dagger(x_2,t_2) \ket}
{\bra \Psi_N(\lambda_1,\cdots,\lambda_N)\rangle_{\epsilon}}\right\},
\end{eqnarray}
where the index $\epsilon=\pm$ represents the boundary conditions.
For the ground state case $T=0$,
the dynamical correlation functions for $N$ state are defined by
the vacuume expectation value of the ground state :
\begin{eqnarray}
\langle \psi(x_1,t_1)\psi^\dagger(x_2,t_2)\rangle_{\epsilon,N,0}
=\displaystyle \frac{\bra \psi(x_1,t_1)\psi^\dagger(x_2,t_2) \ket}
{\bra \Psi_N(\lambda_1,\cdots,\lambda_N)\rangle_{\epsilon}},
\end{eqnarray}
where the spectral parameters 
$(\lambda_1,\cdots,\lambda_N)$ are given by
\begin{eqnarray}
\lambda_j=
\left\{\begin{array}{cc}
\frac{\pi}{L}(j-1), &{\rm for~Neumann},\\
\frac{\pi}{L}j, &{\rm for~Dirichlet}.
\end{array}\right.
\end{eqnarray}
In this paper we are interested in the thermodynamics of the correlation
functions.
For the nonzero temperature $T>0$, the dynamical correlation functions
in the thermodynamic limit are defined by
\begin{eqnarray}
\langle \psi(x_1,t_1)\psi^\dagger(x_2,t_2)\rangle_{\epsilon,T}
=\lim_{N,L \to \infty \atop{\frac{N}{L}=D(T)}}
\langle \psi(x_1,t_1)\psi^\dagger(x_2,t_2)\rangle_{\epsilon,N,T}.
\end{eqnarray}
Here the density $D(T)=\frac{N}{L}$ is given by
\begin{eqnarray}
D(T)=\frac{1}{\pi}
\int_{0}^{\infty}\vartheta(\lambda)d\lambda,
\end{eqnarray}
where the Fermi weight $\vartheta(\lambda)$ is given by
\begin{eqnarray}
\vartheta(\lambda)=\frac{1}{1+\exp\left(\frac{\lambda^2-h}{T}\right)}.
\end{eqnarray}
For the ground state $T=0$, the dynamical correlation function
in the thermodynamic limit is defined by
\begin{eqnarray}
\langle \psi(x_1,t_1)\psi^\dagger(x_2,t_2)\rangle_{\epsilon,0}
=\lim_{N,L \to \infty \atop{\frac{N}{L}=D(0)}}
\langle \psi(x_1,t_1)\psi^\dagger(x_2,t_2)\rangle_{\epsilon,N,0}.
\end{eqnarray}
Here the density $D(0)=\frac{N}{L}$ can be chosen arbitrary.
In this paper we give the Fredholm determinant representations
for the dynamical correlation functions
$\langle \psi(x_1,t_1)\psi^\dagger(x_2,t_2)\rangle_{\epsilon,T},~(T \geq 0
,\epsilon =\pm)$.

\section{Form Factors}
The purpose of
this section is to derive the determinant formulae for the form factors.
First we prepare a lemma.
\begin{lem}~~For the sequences
$\{f_{j,k}\}_{j=1,\cdots,N+1,~k=1,\cdots,N}$ and
$\{g_{j}\}_{j=1,\cdots,N+1}$,
the following holds :
\begin{eqnarray}
\displaystyle \left.\sum_{\sigma \in S_{N+1}}{\rm sgn}\sigma
f_{\sigma(N+1)}\prod_{j=1}^N g_{\sigma(j),j}=
\left(f_{N+1}+\frac{\partial}{\partial \alpha}\right)
\det_{1 \leq j,k \leq N}
\left(g_{j,k}-\alpha f_j \cdot g_{N+1,k}\right)\right|_{\alpha=0}.
\end{eqnarray}\label{det:form}
\end{lem}
{\sl Proof.}~~~Consider the coset decomposition :
\begin{eqnarray}
S_{N+1}=S_N(N+1) \bigcup S_N(N)\cdot(N,N+1) \bigcup \cdots \bigcup
 S_N(1)\cdot(1,N+1),
\nonumber
\end{eqnarray}  
where $S_N(j)$ is permutations of $(1,\cdots,j-1,N+1,j+1,\cdots,N)$.
Rewrite the left side of the equation $(\ref{det:form})$ 
with respect to the coset decomposition :
\begin{eqnarray}
(L.H.S.)&=&\sum_{j=1}^N g_j \sum_{\tau \in S_N(j)}{\rm sgn}(\tau
\cdot (j,N+1))\prod_{k=1 \atop{k \neq j}}^N f_{\tau(k),k}\cdot
f_{\tau(N+1),j}\nonumber \\
&+&g_{N+1}\sum_{\tau \in S_N(N+1)}{\rm sgn}(\tau \cdot(N+1,N+1))
\prod_{k=1}^N f_{\tau(k),k}=(R.H.S.)
\end{eqnarray}
\hfill $\Box$

Now let us consider the field form factor :
\begin{eqnarray}
&&_{\epsilon}\langle \Psi_{N+1}
(\lambda_1,\cdots,\lambda_{N+1})\vert
\psi^\dagger(x)\vert \Psi_N(\mu_1,\cdots,\mu_N)\rangle_{\epsilon}
\nonumber \\
&=&\sqrt{N+1}\int_0^Ldz_1 \cdots \int_0^Ldz_N
\psi_{N+1}^*(z_1,\cdots,z_N,x\vert \lambda_1,\cdots,\lambda_{N+1})
\\
&\times&\psi_{N}(z_1,\cdots,z_N\vert \mu_1,\cdots,\mu_{N})\nonumber\\
&=&\frac{1}{\sqrt{(1+\delta_{\lambda_1,0})(1+\delta_{\mu_1,0})}}
\sum_{\sigma \in S_{N+1}}{\rm sgn}\sigma
(e^{-i\lambda_{\sigma(N+1)}x}+\epsilon
e^{i\lambda_{\sigma(N+1)}x}) \\
&\times&
\prod_{j=1}^N\left\{
\int_0^L dz ~{\rm sgn}(z-x)
(e^{-i\lambda_{\sigma(j)}z}+\epsilon e^{i\lambda_{\sigma(j)}z})
(e^{i\mu_{j}z}+\epsilon e^{-i\mu_{j}z})
\right\}.\nonumber
\end{eqnarray}
To derive the third line, we have used a simple fact :
\begin{eqnarray}
\sum_{\sigma,\tau \in S_N}{\rm sgn}\sigma \tau
\prod_{j=1}^N f_{\sigma(j),\tau(j)}
=N! \sum_{\sigma \in S_N}{\rm sgn}\sigma \prod_{j=1}^N f_{\sigma(j),j}.
\end{eqnarray}
Using lemma \ref{det:form},
we arrive at the determinant formulae for the form factors.

\begin{lem}~~~The field form factors have the determinant formula.
\begin{eqnarray}
&&_{\epsilon}\langle \Psi_{N+1}
(\lambda_1,\cdots,\lambda_{N+1})\vert
\psi^\dagger(x)\vert \Psi_N(\mu_1,\cdots,\mu_N)\rangle_{\epsilon}
\nonumber \\
&=&\left.
\left(C_{\epsilon}(x \vert \lambda_{N+1})+
\frac{\partial}{\partial \alpha}\right)
\det_{1 \leq j,k \leq N}
\left(I_{\epsilon}(x \vert \lambda_j,\mu_k)-
\alpha C_{\epsilon}(x \vert \lambda_j)
I_{\epsilon}(x \vert \lambda_{N+1},\mu_k)\right)\right|_{\alpha=0}.
\end{eqnarray}
Here we have ued 
\begin{eqnarray}
C_{\epsilon}(x \vert \lambda)&=&
\frac{1}{\sqrt{1+\delta_{\lambda,0}}}(e^{-i\lambda x}+
\epsilon e^{i\lambda x}),
\label{def:C}\\
I_{\epsilon}(x \vert \lambda,\mu)&=&
\frac{1}{\sqrt{(1+\delta_{\lambda,0})(1+\delta_{\mu,0})}}\left\{
\frac{4}{\lambda-\mu}\sin(x(\lambda-\mu))\right.\label{def:I}\\
&+&\left.\epsilon\frac{4}{\lambda+\mu}\sin(x(\lambda+\mu))
-2L(\delta_{\lambda,\mu}+\epsilon \delta_{\lambda,0}\delta_{\mu,0})
\right\}.\nonumber
\end{eqnarray}\label{lem:form}
\end{lem}
We can write the field form factors without using integrals.
From the relation $\psi^\dagger(x,t)=e^{iHt}\psi^\dagger(x)e^{-iHt}$,
the dynamical form factors are given by
\begin{eqnarray}
&&_{\epsilon}\langle \Psi_{N+1}
(\lambda_1,\cdots,\lambda_{N+1})\vert
\psi^\dagger(x,t)\vert \Psi_N(\mu_1,\cdots,\mu_N)\rangle_{\epsilon}
\nonumber \\
&=&\exp\left\{it\left(-h+\sum_{j=1}^{N+1}\lambda_j^2
-\sum_{j=1}^N\mu_j^2\right)\right\}~
_{\epsilon}\langle \Psi_{N+1}
(\lambda_1,\cdots,\lambda_{N+1})\vert
\psi^\dagger(x)\vert \Psi_N(\mu_1,\cdots,\mu_N)\rangle_{\epsilon}.
\label{timform}
\end{eqnarray}

\section{Correlation Functions}
The purpose of this section is to derive the Fredholm determinant formulas
of the dynamical correlation functions
$\langle\psi(x_1,t_1)\psi^\dagger(x_2,t_2)\rangle_{\epsilon,T}$.
First we consider the vacuum expectation values of fields operators.
Using the completeness relation (\ref{completeness}),
the vacuum expectation values of two fields are given by
\begin{eqnarray}
&&\frac{_\epsilon\langle \Psi_N(\mu_1,\cdots,\mu_N)\vert
\psi(x_1,t_1)\psi^\dagger(x_2,t_2)
\vert \Psi_N(\mu_1,\cdots,\mu_N)\rangle_\epsilon}
{_\epsilon\langle \Psi_N(\mu_1,\cdots,\mu_N)\vert
\Psi_N(\mu_1,\cdots,\mu_N)\rangle_\epsilon}\nonumber\\
&=&
\sum_{{\rm all}\{\lambda\}_{N+1}}
\frac{_\epsilon\langle \Psi_N(\{\mu\})\vert
\psi(x_1,t_1)\vert \Psi_{N+1}(\{\lambda\})\rangle_{\epsilon}
~_{\epsilon}\langle \Psi_{N+1}(\{\lambda\}) \vert 
\psi^\dagger(x_2,t_2)\vert \Psi_N(\{\mu\})\rangle_\epsilon}
{_{\epsilon}\langle \Psi_N(\{\mu\})\vert
\Psi_N(\{\mu\})\rangle_{\epsilon}
~_{\epsilon}\langle \Psi_{N+1}(\{\lambda\})\vert
\Psi_{N+1}(\{\lambda\})\rangle_{\epsilon}}.
\end{eqnarray}
Using the equation (\ref{timform}) and the following relations :
\begin{eqnarray}
&&_{\epsilon}\langle \Psi_N(\{\mu\})\vert\psi(x_1,t_1)
\vert \Psi_{N+1}(\{\lambda\})\rangle_{\epsilon}
=_{\epsilon}\langle \Psi_{N+1}(\{\lambda\})\vert\psi^\dagger(x_1,t_1)
\vert \Psi_{N}(\{\mu\})\rangle_{\epsilon}^{*},\\
&&_{\epsilon}\langle \Psi_N(\{\lambda\})\vert
\Psi_N(\{\lambda\})\rangle_{\epsilon}
=(2L)^N,
\end{eqnarray}
we obtain
\begin{eqnarray}
&&\frac{1}{(N+1)!}
\left(\frac{1}{2L}\right)^{2N+1}
e^{-i(t_2-t_1)(h+\sum_{j=1}^N\mu_j^2)}
\sum_{\lambda_1 \in \frac{\pi}{L}{\bf N}}\cdots
\sum_{\lambda_{N+1} \in \frac{\pi}{L}{\bf N}}\\
&\times&
e^{i(t_2-t_1)\sum_{j=1}^{N+1}\lambda_j^2}
~_{\epsilon}\langle \Psi_{N+1}(\{\lambda\})\vert\psi^\dagger(x_1)
\vert \Psi_{N}(\{\mu\})\rangle_{\epsilon}^{*}
~_{\epsilon}\langle \Psi_{N+1}(\{\lambda\})\vert\psi^\dagger(x_2)
\vert \Psi_{N}(\{\mu\})\rangle_{\epsilon}.\nonumber
\end{eqnarray}
The translation invariance of time holds.
\begin{eqnarray}
_\epsilon\langle \Psi_N(\{\mu\})\vert
\psi(x_1,t_1)\psi^\dagger(x_2,t_2)
\vert \Psi_N(\{\mu\})\rangle_\epsilon
=
_\epsilon\langle \Psi_N(\{\mu\})\vert
\psi(x_1,0)\psi^\dagger(x_2,t_2-t_1)
\vert \Psi_N(\{\mu\})\rangle_\epsilon.
\end{eqnarray}
In the sequel we set the abberiviation $t=t_2-t_1$.
Remember a following simple fact.
\begin{lem}~~For sequences $\{f_{j_1,\cdots,j_n}\}
_{j_1,\cdots,j_n \in I},
\{g_{j_1,\cdots,j_n}\}
_{j_1,\cdots,j_n \in I},
~(I : {\rm some~index~set})$,
the following holds.
\begin{eqnarray}
\sum_{j_1,\cdots,j_n \in I}
({\rm Sym}~f)_{j_1,\cdots,j_n}
({\rm Sym}~g)_{j_1,\cdots,j_n}
=\sum_{j_1,\cdots,j_n \in I}
f_{j_1,\cdots,j_n}~
({\rm Sym}~g)_{j_1,\cdots,j_n}.
\end{eqnarray}
Here we have used 
\begin{eqnarray}
({\rm Sym}~f)_{j_1,\cdots,j_n}=\frac{1}{n!}
\sum_{\sigma \in S_n}f_{j_{\sigma(1)},\cdots,j_{\sigma(n)}}.
\end{eqnarray}\label{lem:sym}
\end{lem}
The form factors have the determinant formulae in lemma 
\ref{lem:form} and
\begin{eqnarray}
_{\epsilon}\langle \Psi_{N+1}(\{\lambda\})\vert\psi^\dagger(x)
\vert \Psi_{N}(\{\mu\})\rangle_{\epsilon}
=\sum_{\sigma \in S_{N+1}}
C_{\epsilon}(x \vert \lambda_{\sigma(N+1)})
\prod_{j=1}^N I_{\epsilon}(x \vert \lambda_{\sigma(j)},\mu_j),
\end{eqnarray}
Therefore lemma \ref{lem:sym}, we obtain
\begin{eqnarray}&&
e^{-it(h+\sum_{j=1}^N\mu_j^2)}\left(\frac{1}{2L}\right)^{2N+1}
\sum_{\lambda_1 \in \frac{\pi}{L}{\bf N}}
\cdots
\sum_{\lambda_{N+1} \in \frac{\pi}{L}{\bf N}}
\left(e^{it\lambda_{N+1}^2}C_{\epsilon}^*(x_1 \vert \lambda_{N+1})
C_{\epsilon}(x_2 \vert \lambda_{N+1})
+\frac{\partial}{\partial \alpha}\right)\nonumber\\
&\times&\left.\det_{1 \leq j,k \leq N}\left(
e^{it\lambda_j^2}I_{\epsilon}(x_1 \vert \lambda_j, \mu_k)
I_{\epsilon}(x_2 \vert \lambda_j, \mu_j)-\alpha
e^{it\lambda_j^2}C^*_{\epsilon}(x_1 \vert \lambda_j)
I_{\epsilon}(x_2 \vert \lambda_j, \mu_j)\right. \right.\\
&\times&\left. \left.
e^{it\lambda_{N+1}^2}C_{\epsilon}(x_2 \vert \lambda_{N+1})
I_{\epsilon}(x_1 \vert \lambda_{N+1}, \mu_k)
\right)\right|_{\alpha=0}.\nonumber
\end{eqnarray}
The $j$ th line of the above matrix only depends on $\lambda_j$
not on $\lambda_k,(k \neq j)$, therefore we can insert the summations
$\sum_{\lambda_1}\cdots \sum_{\lambda_{N+1}}$
 into the matrix. Now we arrive at the following.

\begin{prop}~~~The vacuum expectation values of two fields have
the determinant formulas.
\begin{eqnarray}
&&\frac{_\epsilon\langle \Psi_N(\mu_1,\cdots,\mu_N)\vert
\psi(x_1,0)\psi^\dagger(x_2,t)
\vert \Psi_N(\mu_1,\cdots,\mu_N)\rangle_\epsilon}
{_\epsilon\langle \Psi_N(\mu_1,\cdots,\mu_N)\vert
\Psi_N(\mu_1,\cdots,\mu_N)\rangle_\epsilon}\nonumber\\
&=&
e^{-ith} 
\left(
\frac{1}{2L}\sum_{s \in \frac{\pi}{L}{\bf N}}
\epsilon e^{it s^2}C_{\epsilon}(x_1 \vert s)
C_{\epsilon}(x_2 \vert s)
+\frac{\partial}{\partial \alpha}\right)
\left.\det_{1 \leq j,k \leq N}\left(
\left(\frac{1}{2L}\right)^2
\sum_{s \in \frac{\pi}{L}{\bf N}}
e^{it s^2}J_{\epsilon}(x_1 \vert s, \mu_k)
J_{\epsilon}(x_2 \vert s, \mu_j)\right.\right.\nonumber\\
&-&\left.\left.\alpha \epsilon \frac{1}{2L}
\left(\frac{1}{2L}
\sum_{s \in \frac{\pi}{L}{\bf N}}
e^{it s^2}C_{\epsilon}(x_1 \vert s)
J_{\epsilon}(x_2 \vert s, \mu_j)\right)\right. \right.
\left. \left.
\left(\frac{1}{2L}
\sum_{s \in \frac{\pi}{L}{\bf N}}
e^{it s^2}C_{\epsilon}(x_2 \vert s)
J_{\epsilon}(x_1 \vert s, \mu_k)\right)
\right)\right|_{\alpha=0}.
\end{eqnarray}
Here we have used 
\begin{eqnarray}
J_{\epsilon}(x\vert s ,\mu)=e^{-\frac{1}{2}it\mu^2}
I_{\epsilon}(x\vert s,\mu).
\end{eqnarray}
and functions $C_{\epsilon}(x \vert s)$
 and $I_{\epsilon}(x\vert s, \mu)$
are defind in (\ref{def:C}) and (\ref{def:I}), respectively.
\end{prop}
The size of the above matrix depends on the state number $N$,
however, the element of the matrix does not depend on $N$.
By calculations, we obtan
\begin{eqnarray}
&&\frac{1}{2L}\sum_{s \in \frac{\pi}{L}{\bf N}}
\epsilon e^{it s^2}C_{\epsilon}(x_1 \vert s)
C_{\epsilon}(x_2 \vert s)
=\frac{1}{2L}\sum_{s \in \frac{\pi}{L}{\bf Z}}
e^{it s^2-is(x_1-x_2)}
+\epsilon
\frac{1}{2L}\sum_{s \in \frac{\pi}{L}{\bf Z}}
e^{it s^2-is(x_1+x_2)},\\
&&\frac{1}{2L}
\sum_{s \in \frac{\pi}{L}{\bf N}}e^{it s^2}
C_{\epsilon}(x_1 \vert s)
J_{\epsilon}(x_2 \vert s, \mu)\\
&=&\frac{e^{-\frac{1}{2}it\mu^2}}{\sqrt{1+\delta_{\mu,0}}}
\left[
\left\{
\frac{2}{\pi}\left(\frac{\pi}{L}\right)
\sum_{s \in \frac{\pi}{L}{\bf Z}}
\frac{e^{it s^2-is x_1}}
{s-\mu}\sin(x_2(s-\mu))
-e^{it\mu^2-i\mu x_1}\right\}+\epsilon
\{~~\mu \leftrightarrow (-\mu)~~\}\right],\nonumber 
\end{eqnarray}
and
\begin{eqnarray}
&&\left(\frac{1}{2L}\right)^2
\sum_{s \in \frac{\pi}{L}{\bf N}}
e^{it s^2}J_{\epsilon}(x_1 \vert s, \mu)
J_{\epsilon}(x_2 \vert s, \lambda) \\
&=&
\delta_{\lambda,\mu}-\frac{2}{\pi}\left(\frac{\pi}{L}\right)
\frac{e^{-\frac{1}{2}it(\lambda^2+\mu^2)}}
{\sqrt{(1+\delta_{\lambda,0})(1+\delta_{\mu,0})}}
\left[\frac{1}{\lambda-\mu}\left\{
e^{it\lambda^2}\sin(x_1(\lambda-\mu))+
e^{it\mu^2}\sin(x_2(\lambda-\mu))\right.\right.\nonumber \\
&-&\left.\left.\frac{2}{\pi}\left(\frac{\pi}{L}\right)
\sum_{s \in \frac{\pi}{L}{\bf Z}}
e^{it s^2}\left(\frac{1}{s-\lambda}-\frac{1}{s-\mu}\right)
\sin((s-\mu)x_1)\sin((s-\lambda)x_2)\right\}
+\epsilon \frac{1}{\lambda+\mu}\{~\mu \leftrightarrow (-\mu)\}\right].
\nonumber
\end{eqnarray}
Therefore we can take the thermodynamic limits.
Let us set 
\begin{eqnarray}
\tau(s,x)&=&its^2-ixs,\\
G(x)&=&\int_{-\infty}^{\infty}e^{\tau(s,x)}ds.
\end{eqnarray}
The following are our main results.

\begin{thm}~~~In the thermodynamic limit
$N,L \to \infty,~{\rm such~that}~\frac{N}{L}=D$,
the ground state dynamical correlation functions
have the Fredholm determinant representations.
\begin{eqnarray}
&&\langle\psi(x_1,0)\psi^\dagger(x_2,t_2)\rangle_{\epsilon,0}\nonumber \\
&=&\frac{e^{-iht}}{2\pi}
\left(G(x_1-x_2)+\epsilon G(x_1+x_2)+\frac{\partial}{\partial \alpha}\right)
\left.\det\left(1-\frac{2}{\pi}
\widehat{V}_{\epsilon}-\alpha \widehat{W}_{\epsilon}\right)\right|_{\alpha=0}.
\end{eqnarray}
Here the integral operators $\widehat{V}_{\epsilon}$ and
$\widehat{W}_{\epsilon}$ are defined by
\begin{eqnarray}
(\widehat{V}_{\epsilon}f)(\lambda)
=\int_{0}^q {V}_{\epsilon}(\lambda,\mu)f(\mu)d \mu,~~
(\widehat{W}_{\epsilon}f)(\lambda)
=\int_{0}^q {W}_{\epsilon}(\lambda,\mu)f(\mu)d \mu,
\end{eqnarray}
where the Fermi sphere $q=\pi D$ and the integral kernel are given by
Neumann or Dirichlet sum :
\begin{eqnarray}
V_{\epsilon}(\lambda,\mu)&=&e^{-\frac{1}{2}it(\lambda^2+\mu^2)}
\left\{\frac{1}{\lambda-\mu}L(\lambda,\mu)
+\epsilon \frac{1}{\lambda+\mu}L(\lambda,-\mu)\right\},\\
W_{\epsilon}(\lambda,\mu)&=&\epsilon
e^{-\frac{1}{2}it(\lambda^2+\mu^2)}
\left\{P(\lambda \vert x_1,x_2)+\epsilon P(-\lambda \vert x_1,x_2)\right\}
\left\{P(\mu \vert x_2,x_1)+\epsilon P(-\mu \vert x_2,x_1)\right\}.
\end{eqnarray}
Here we have used
\begin{eqnarray}
L(\lambda,\mu)&=&e^{it\lambda^2}\sin(x_1(\lambda-\mu))+
e^{it\mu^2}\sin(x_2(\lambda-\mu))\\
&~&
+\frac{2}{\pi}\int_{-\infty}^{\infty}
\left(\frac{1}{s-\mu}-\frac{1}{s-\lambda}\right)e^{its^2}
\sin((s-\mu)x_1)\sin((s-\lambda)x_2)ds.\nonumber
\\
P(\lambda \vert x_1,x_2)&=&e^{\tau(\lambda,x_1)}-\frac{2}{\pi}
\int_{-\infty}^{\infty}\frac{1}{s-\lambda}e^{\tau(s,x_1)}\sin(x_2(s-\lambda))
ds.
\end{eqnarray}
\end{thm}
We have suceeded to write the integral kernel by elementary functios :
\begin{eqnarray}
\int_{-\infty}^{\infty}\frac{e^{\tau(s,y)}}{s-y}ds,
\end{eqnarray}
and trigonometric functions.
Next we study the finite temperatute thermodynamics.
By statistical mechanics arguments,
at temperature $T>0$,
the thermodynamic equilibrium distribution of the spectral parameters
is given by the Fermi weight $\vartheta(\lambda)$ :
\begin{eqnarray}
\lim ~\left(\frac{\pi}{L}\right)
\frac{1}{\lambda_{j+1}-\lambda_j}=
\vartheta(\lambda_j),~~{\rm where}~~~
\vartheta(\lambda)=\frac{1}{1+\exp\left(\frac{\lambda^2-h}{T}\right)}.
\label{Fermi}
\end{eqnarray}
Therefore the density is given by
\begin{eqnarray}
D(T)=\frac{N}{L}=\frac{1}{\pi}\int^{\infty}_0 \vartheta(\lambda)d\lambda.
\end{eqnarray}
We arrive at the following theorem.

\begin{thm}~~~In the thermodynamic limit :
$N,L \to \infty,{\rm such~that~}\frac{N}{L}=D(T)$,
the finite temperature dynamical correlation functions
have the Fredholm determinant representations.
\begin{eqnarray}
&&\langle\psi(x_1,0)\psi^\dagger(x_2,t_2)\rangle_{\epsilon,T}\nonumber \\
&=&\frac{e^{-iht}}{2\pi}
\left(G(x_1-x_2)+\epsilon G(x_1+x_2)+\frac{\partial}{\partial \alpha}\right)
\left.\det\left(1-\frac{2}{\pi}
\widehat{V}_{\epsilon,T}-\alpha 
\widehat{W}_{\epsilon,T}\right)\right|_{\alpha=0}.
\end{eqnarray}
Here the temperature $T>0$ and
the integral operators $\widehat{V}_{\epsilon,T}$ and
$\widehat{W}_{\epsilon,T}$ are defined by
\begin{eqnarray}
(\widehat{V}_{\epsilon,T}f)(\lambda)
=\int_{0}^{\infty} {V}_{\epsilon}(\lambda,\mu)
\vartheta(\mu)f(\mu)d \mu,~~
(\widehat{W}_{\epsilon,T}f)(\lambda)
=\int_{0}^{\infty} {W}_{\epsilon}(\lambda,\mu)
\vartheta(\mu)f(\mu)d \mu,
\end{eqnarray}
where the Fermi weight $\vartheta(\lambda)$ 
is given in (\ref{Fermi}).
\end{thm}

\section{Degenerate Case}
The purpose of this section is to derive the Fredholm minor determinant
representations for finite-temperature fields correlation functions :
\begin{eqnarray}
\langle \psi(x_1)\psi^\dagger(x_2)\rangle_{\epsilon,T}.
\end{eqnarray}
Especially our Fredholm minor determinant representations
coincide with the one which has been obtained in \cite{K}.
When we take the limit $t \to 0$, the following simplifications occur :
\begin{eqnarray}
G(x)\to 0,~~
L(\lambda,\mu)\to \sin(x_1(\lambda-\mu))+\sin(x_2(\lambda-\mu)),~~
P(\lambda \vert x_1,x_2)\to e^{-i x_1 \lambda}.
\end{eqnarray}
Therefore we obtain
\begin{eqnarray}
\langle \psi(x_1)\psi(x_2) \rangle_{\epsilon,T}
&=&\left.\frac{1}{2\pi}
\left(\frac{\partial}{\partial \alpha}\right)
\det\left(1-\frac{2}{\pi}\widehat{\tilde{V}}_{\epsilon,T}
-\alpha \widehat{\tilde{W}}^{(x_1,x_2)}_{\epsilon,T}\right)
\right|_{\alpha=0}\\
&=&-\frac{1}{2\pi}
\det\left(1-\frac{2}{\pi}\widehat{\tilde{V}}_{\epsilon,T}\right)
{\rm Tr}\left[\left(1-\frac{2}{\pi}
\widehat{\tilde{V}}_{\epsilon,T}\right)^{-1}
\widehat{\tilde{W}}_{\epsilon,T}^{(x_1,x_2)}\right],
\end{eqnarray}
where the integral operators are given by
\begin{eqnarray}
\left(\widehat{\tilde{V}}_{\epsilon,T}f\right)(\lambda)=\int_0^{\infty}
\tilde{V}_{\epsilon,T}(\lambda,\mu)f(\mu)d\mu,~~
\left(\widehat{\tilde{W}}_{\epsilon,T}^{(\xi,\eta)}
f\right)(\lambda)=\int_0^{\infty}
\tilde{W}_{\epsilon,T}^{(\xi,\eta)}(\lambda,\mu)f(\mu)d\mu.
\end{eqnarray}
and the integral kernels are given by
\begin{eqnarray}
\tilde{V}_{\epsilon,T}(\lambda,\mu)
&=&\sqrt{\vartheta(\lambda)}\left[
\frac{1}{\lambda-\mu}\left\{\sin(x_1(\lambda-\mu))+
\sin(x_2(\lambda-\mu))\right\}\right.\\
&+&\left.\epsilon \frac{1}{\lambda+\mu}\left\{\sin(x_1(\lambda+\mu))+
\sin(x_2(\lambda+\mu))\right\}\right]\sqrt{\vartheta(\mu)},\nonumber\\
\tilde{W}_{\epsilon,T}^{(\xi,\eta)}
(\lambda,\mu)&=&\sqrt{\vartheta(\lambda)}
\epsilon(e^{i \xi \lambda}+\epsilon e^{-i \xi \lambda})
(e^{i \eta \mu}+\epsilon e^{-i \eta \mu})\sqrt{\vartheta(\mu)}.
\end{eqnarray}
Pay attention to the Fourier transforms :
\begin{eqnarray}
f(\lambda)=\frac{1}{2 \pi \sqrt{\vartheta(\lambda)}}
\int_{-\infty}^{\infty} d\xi e^{i\lambda \xi}\varphi(\xi),~~
\varphi(\xi)=\int_{-\infty}^{\infty} d\lambda \sqrt{\vartheta(\lambda)}
e^{-i\lambda \xi}f(\lambda).
\end{eqnarray}
The following identity holds for functions 
$f_{\epsilon}(\epsilon \lambda)=\epsilon f(\epsilon \lambda)$.
\begin{eqnarray}
&&\int_0^{\infty}d\mu f_{\epsilon}(\mu)
\sqrt{\vartheta(\lambda)\vartheta(\mu)}
\left\{\frac{1}{\lambda-\mu}\sin(x(\lambda-\mu))+
\epsilon \frac{1}{\lambda+\mu}\sin(x(\lambda+\mu))\right\}\nonumber \\
&=&
\frac{1}{2\pi\sqrt{\vartheta(\lambda)}}
\int_{-\infty}^{\infty}d\xi e^{i\lambda \xi}
\left(\int_0^x d\xi' \theta_{\epsilon,T}(\xi',\xi)\right)
\int_{-\infty}^{\infty}d\mu
f_{\epsilon}(\mu)\sqrt{\vartheta(\mu)}e^{-i\xi'\mu},
\end{eqnarray}
where
\begin{eqnarray}
\theta_{\epsilon,T}(\xi,\eta)
=\int_0^{\infty}\vartheta(\nu)
\left\{\cos((\xi-\eta)\nu)+\epsilon \cos((\xi+\eta)\nu)\right\}d\nu.
\end{eqnarray}
Therefore we arrive at 
\begin{eqnarray}
\det\left(1-\frac{2}{\pi}\widehat{\tilde{V}}_{\epsilon,T}\right)
=\det\left(1-\frac{2}{\pi}\left(\hat{\theta}_{\epsilon,T}^{(-x_1,x_2)}
\right)\right),
\end{eqnarray}
where the integral operator $\hat{\theta}_{\epsilon,T}^{(y_1,y_2)}$
is defined by
\begin{eqnarray}
\left(\hat{\theta}_{\epsilon,T}^{(y_1,y_2)}f\right)(\xi)
=\int_{y_1}^{y_2}\theta_{\epsilon,T}(\xi,\xi')f(\xi')d\xi'.
\label{def:theta}
\end{eqnarray}
Let us set
\begin{eqnarray}
\Delta_{\epsilon}(\xi,\eta)=\frac{\det
\left(1-\frac{2}{\pi}\left(\hat{\theta}_{\epsilon,T}^{(-x_1,x_2)}\right)\left|
\begin{array}{c}\eta\\
\xi
\end{array}\right.\right)}
{\det\left(1-\frac{2}{\pi}\left(\hat{\theta}_{\epsilon,T}^{(-x_1,x_2)}
\right)\right)}.
\end{eqnarray}
Here we have used the following notation of the Fredholm minor determinants :
\begin{eqnarray}
\det
\left(1-\lambda \widehat{K}_I\left|
\begin{array}{ccc}\xi_1 & \cdots & \xi_r\\
\eta_1 & \cdots &\eta_r
\end{array}\right.\right)
=
\sum_{n=0}^{\infty}\frac{(-\lambda)^{n+r}}{n!}
\int_I d\lambda_1 \cdots 
\int_I d\lambda_n K_{n+r}\left(\begin{array}{cccccc}\xi_1 & \cdots & \xi_r 
~\lambda_1 &\cdots &\lambda_n\\
\eta_1 & \cdots &\eta_r
~\lambda_1 &\cdots &\lambda_n
\end{array}\right),
\end{eqnarray}
where we have used
\begin{eqnarray}
K_{m}\left(\begin{array}{ccc}\xi_1 & \cdots & \xi_m \\
\eta_1 & \cdots &\eta_m
\end{array}\right)
=\det_{1 \leq j,k \leq m}
\left(K(\xi_j,\eta_k)\right).
\end{eqnarray}
The integral operator $\widehat{K}_I$ is defined by using
the integral kernel $K(\lambda,\mu)$ and the integral interval $I$ :
\begin{eqnarray}
(\widehat{K}_I f)(\lambda)=\int_I K(\lambda,\mu)f(\mu) d\mu.
\end{eqnarray}
From definition,
the function $\Delta_{\epsilon}(\xi,\eta)$ satisfies the integral equation :
\begin{eqnarray}
\Delta_{\epsilon}(\xi,\eta)-\frac{2}{\pi}
\int_{-x_1}^{x_2}\theta_{\epsilon,T}(\xi,\xi')
\Delta_{\epsilon}(\xi',\eta)d\xi'=
-\frac{2}{\pi}\theta_{\epsilon,T}(\xi,\eta).
\end{eqnarray}
Let us take the Fourier transforms of this integral equation.
\begin{eqnarray}
&&
\frac{1}{2\pi \sqrt{\vartheta(\lambda)}}
\int_{-\infty}^{\infty}d\xi e^{i\lambda \xi}\Delta_{\epsilon}(\xi,\eta)\\
&-&\frac{2}{\pi}\int_{-x_1}^{x_2}d\mu \tilde{V}_{\epsilon,T}
(\lambda,\mu)
\frac{1}{2\pi \sqrt{\vartheta(\mu)}}
\int_{-\infty}^{\infty}d\xi'e^{i\mu \xi'}\Delta_{\epsilon}(\xi',\eta)
=-\frac{1}{\pi}\sqrt{\vartheta(\lambda)}(e^{i\lambda \eta}+
\epsilon e^{-i\lambda \eta}).\nonumber
\end{eqnarray}
Therefore we know
\begin{eqnarray}
\Delta_{\epsilon}(\xi,\eta)=
-\frac{1}{\pi}{\rm Tr}\left[
\left(1-\frac{2}{\pi}\widehat{\tilde{V}}_{\epsilon,T}\right)^{-1}
\widehat{\tilde{W}}_{\epsilon,T}^{(\xi,\eta)}\right].
\end{eqnarray}
Now we arrive at
\begin{cor}~~The finite temperature field correlation functions 
have the Fredholm minor determinants representations.
\begin{eqnarray}
\langle \psi(x_1)\psi^{\dagger}(x_2)\rangle_{\epsilon,T}
=\frac{1}{2}\det\left(1-\frac{2}{\pi}\widehat{\theta}^{(x_1,x_2)}_{\epsilon,T}
\left|\begin{array}{c}
x_2 \\ x_1
\end{array}\right.
\right),
\end{eqnarray}
where the integral operator 
$\widehat{\theta}^{(x_1,x_2)}_{\epsilon,T}$ is defined in (\ref{def:theta}).
\end{cor}
In much the same way as with finite temperature case,
we obtain the Fredholm determinant representations for the ground
state correlation functions.
\begin{cor}~~~The ground state correlation functions have the
Fredholm minor determinant formulae.
\begin{eqnarray}
\langle \psi(x_1)\psi^{\dagger}(x_2)\rangle_{\epsilon,0}
=\frac{1}{2}\det\left(1-\frac{2}{\pi}\widehat{K}^{(x_1,x_2)}_{\epsilon}
\left|\begin{array}{c}
x_2 \\ x_1
\end{array}\right.
\right).
\end{eqnarray}
Here the integral operator is defined by
\begin{eqnarray}
\left(\hat{K}_{\epsilon}^{(x_1,x_2)}f\right)(\xi)
=\int_{x_1}^{x_2}K_{\epsilon}(\xi,\xi')f(\xi')d\xi',
\end{eqnarray}
where
\begin{eqnarray}
K_{\epsilon}(\xi,\eta)=\frac{\sin D(\xi-\eta)}{\xi-\eta}
+\epsilon
\frac{\sin D(\xi+\eta)}{\xi+\eta}.
\end{eqnarray}
Here the density $D=\frac{N}{L}$.
\end{cor}
This result coincides with the one which has been obtained 
by the help of fermions \cite{K}.

~

{\sl Acknowledgements}~~~
This work is partly supported by the Japan Society for Promotion of Science.

~

\end{document}